# Catastrophic thermal destabilization of two-dimensional close-packed emulsions due to synchronized coalescence initiation

**Samira Abedi, Chau-Chyun Chen* and Siva A. Vanapalli***

[a] Department of Chemical Engineering, Texas Tech University, Lubbock, TX 79409-3121, USA

*Corresponding author. Email address: siva.vanapalli@ttu.edu; chauchyun.chen@ttu.edu

**Abstract**

The mechanisms for phase separation in highly concentrated emulsions when subjected to a thermal phase transition remain to be elucidated. Here, we create a hexagonally close-packed monodisperse emulsion in 2D and show that during a cool-heat cycle, the emulsion fully destabilizes akin to phase separation. The mechanism for this catastrophic destabilization is found to be spontaneous coalescence initiation that synchronously occurs between every solidified droplet and its neighbors. This synchronous coalescence initiation establishes system-wide network connectivity in the emulsion causing large-scale destabilization. This system-wide coalescence initiation is found to be insensitive to droplet size and surfactant type, but dependent on network connectivity and crystal content of individual droplets.

Concentrated emulsions with crystallizable oil drops, when cooled and heated, can undergo destabilization with changes in droplet size distribution, which in some instances, leads to phase separation [1-5]. This thermal treatment highlights that emulsion destabilization can be influenced by dispersed phase crystallization. Studies show that one of the key mechanisms that induce the emulsion destabilization process is partial coalescence [6-10]. Crystalline droplets contacting each other partially coalesce because the elastic stress of the crystal network dominates the interfacial stress preventing total coalescence of the doublets or multibody droplet aggregates [11]. During heating, these partially-coalesced structures relax shape to form larger spherical droplets, thereby destabilizing emulsions [1, 12]. Tuning the degree of partial coalescence is a functional route to controlling the properties of foods [13-18], consumer products [4, 19], and phase change materials [20] and more recently has been exploited to sculpt novel anisotropic structures [21, 22].

Despite decades of investigation on thermal destabilization of emulsions, details of the mechanisms that connect droplet-level properties to system-level destabilization remain unknown. Studies investigated the effects of droplet-level properties such as dispersed phase volume fraction [5], crystal content [23-25], droplet size [4, 6, 9, 26], and surfactant [4]. These studies were performed in bulk emulsions with indirect techniques, without visualization of the destabilization process, providing at best correlative results on droplet-level factors affecting emulsion destabilization [27]. Thus, there has been a need to develop new approaches that allow direct observation of how individual droplets interact with each other to produce system-wide effects.

In one recent approach, a pair or a triplet of isothermal semi-crystalline droplets were brought together by micromanipulators to force partial coalescence [11, 28-30]. With a pair of droplets, the results showed that droplets may or may not coalesce depending on the crystal content, highlighting the relative importance of elastic and interfacial stresses. The studies with triplets showed that depending on the angle between droplets, the triplet could restructure via expansion of the menisci [28]. This capillarity or meniscus-driven restructuring leads to more close-packed structures, which at the densest packing arrange droplets at 60° bond angles [28, 31]. These insights are beginning to address phenomena at the level of droplet-clusters but still do not reach the scale of a large ensemble of droplets representative of real emulsions.

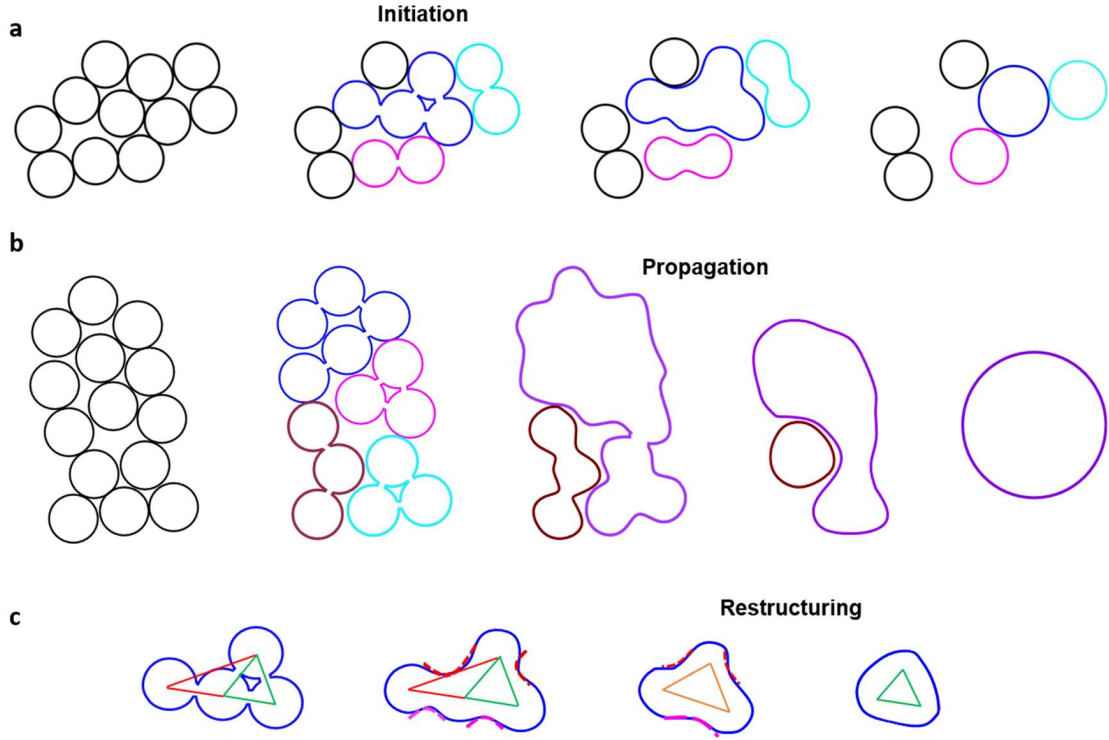

**Fig. 1. Mechanisms of emulsion destabilization in disordered ensembles of partially crystalline droplets.** (**a**) Coalescence initiation – During the heating cycle, the crystalline droplets form spontaneous bridges producing small clusters. As the melting proceeds (from left to right), the clusters relax shape to form larger spherical drops. (**b**) Coalescence propagation - As the melting proceeds, the shape-relaxing clusters contact each other. When the interfacial curvatures match, the clusters combine to form a large liquid spherical droplet. (**c**) Capillary restructuring. The shape relaxation of a 4-droplet aggregate through meniscus-driven restructuring. As melting proceeds, the angles between the droplets evolve to the most compact form of 60°.

Recently, we introduced an approach where the thermal destabilization of an ensemble of monodisperse oil-in-water emulsions is visualized in two-dimensions by geometric confinement in a rectangular chamber [32]. Using this approach, we showed that in *disordered* emulsions subjected to a cool-heat cycle, the destabilization could be understood as a two-step process involving coalescence initiation and coalescence propagation (Fig. 1a,b). As shown in Fig. 1a, coalescence initiation involves spontaneous bridging of droplets producing small clusters (Fig. 1a). During coalescence propagation, these small clusters were found to merge producing large-scale structures (Fig. 1b). The general picture that emerged from this study is that starting from individual droplets, the probability of coalescence propagation depended on the number of contacts with a given droplet, the directionality of these contacts, and capillary-driven restructuring (Fig. 1c). Thus, in addition to physiochemical properties of individual droplets, geometric and kinetic barriers are present that can influence the degree of thermal destabilization in emulsions.

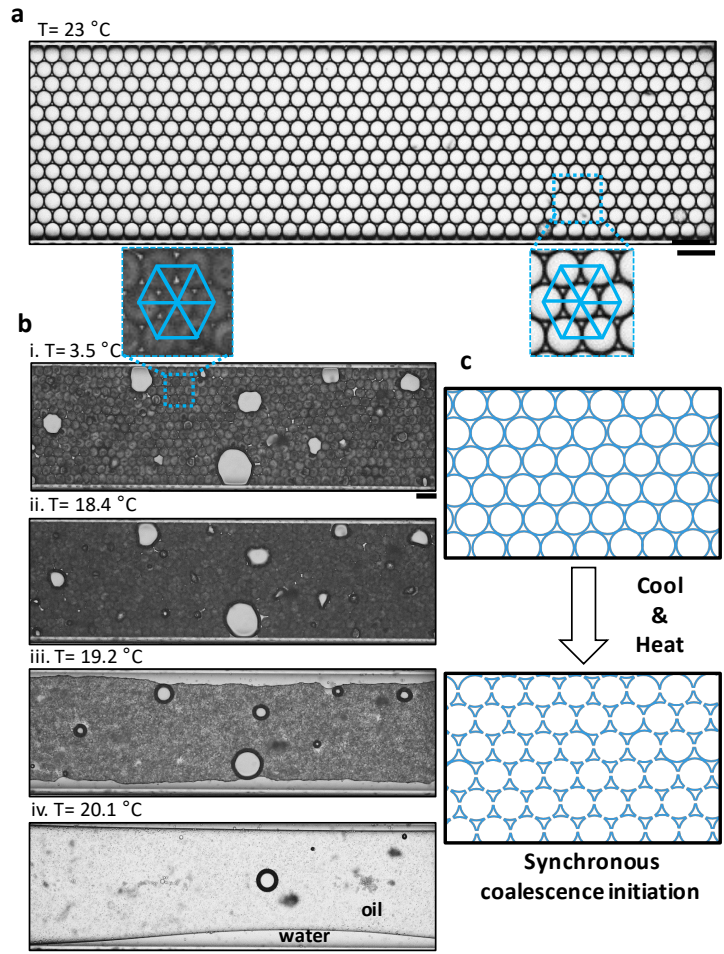

**Fig. 2. Catastrophic thermal destabilization of a 2D hexagonal emulsion. (a)** Image showing the 2D hexagonal droplet array for studying thermal destabilization of emulsions. The inset highlights a hexagonal cluster where the droplet in the center has 6 neighbors with 60° bond angles. (**b**) Image sequence of the destabilization and phase separation of a n-hexadecane 2D emulsion upon cooling and heating at 1°C/min. The droplet size is 40 μm and stabilized by 2 wt% sodium dodecyl sulfate. (i) The emulsion is fully crystallized at 3.5°C. Bubbles (white patches) nucleate due to the thermal contraction of the emulsion. The inset shows a hexagonal solid cluster with the droplets not undergoing partial coalescence. (ii-iv) During heating from 3.5 °C to 20.1 °C, coalescence initiates between the droplets synchronously, and the emulsion destabilizes into two separate oil and water phases. (**c**) Schematic representation of the mechanism causing catastrophic destabilization of the 2D emulsion. During the heating cycle, synchronous coalescence is initiated making bridges between every individual droplet and their six neighbors. This process creates a network of partially coalesced droplets which phase separates. Scale bar denotes 100 μm.

Given that the emulsion destabilization process not only depends on the properties of individual droplets but also on the number and orientation of contacts, we sought to develop an approach where some of these factors are tightly controlled. In this study, we create a two-dimensional (2D) emulsion configuration consisting of hexagonally close-packed droplet arrays. This symmetric arrangement allows us to (i) controllably vary individual droplet properties, and (ii) eliminate cluster-level dynamics. As shown in Fig. 2a, the hexagonal packing ensures each droplet has the same number of six neighbors in contact, and the directionality of connections between each droplet is also the same since the bond angles are 60°. Using this geometrically-constrained configuration, we study how individual droplet properties such as size, surfactant, and crystal content influence system-wide destabilization.

Figure 2a,b show the representative images from the thermal cycling of a 40 μm emulsion stabilized by 2wt% sodium dodecyl sulfate (SDS). The dispersed phase is *n*-hexadecane with a bulk melting point of $T_m$ = 18.2 °C. The emulsion is confined in a chamber of height 50 μm and the calculated volume fraction φ is 0.4. Initially, the emulsion consists of hexagonally packed liquid droplets at room temperature (Fig. 2a), which are cooled at a linear cooling rate of 1°C/min until the entire emulsion is crystallized at 3.5 °C. During cooling, droplets are observed to undergo collective nucleation dynamics, which we discussed in detail recently [33]. As shown in Fig. 2b-i, the crystalline droplets become non-spherical and no partial coalescence was observed during cooling [33]. Additionally, the thermal contraction of the emulsion reduces pressure and induces growth of nucleated bubbles in the constant-volume geometry (see Supplementary Movie SM1). Details of the experimental methodology are provided in Ref [32, 33].

When the emulsion is heated, the dark boundaries of the droplets start to become blurred at 18.4 °C (Fig. 2b-ii) indicating coalescence initiation between the melting droplets. Strikingly, this coalescence initiation occurs at the same time in all the droplets in the array triggering destabilization of the emulsion. In Fig. 2b-iii, we observe that at 19.2 °C, all the boundaries between the droplets disappear, and the emulsion transforms into one large partially crystalline lump. Finally, upon further heating to 20.1 °C, the original emulsion separates into two distinct phases of oil and water (Fig. 2b-iv, also see SM 1).

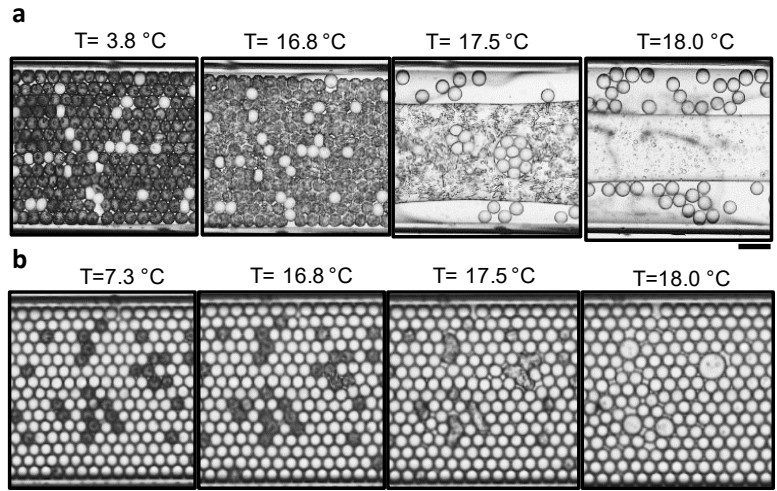

**Fig. 3. Impact of solid-drop connectivity on triggering synchronous coalescence initiation.** The *n*-hexadecane emulsion stabilized with 2 wt% SDS was cooled such that all the droplets do not fully crystallize (**a**) Emulsion was cooled to 3.8°C producing a solidified droplet fraction of $\approx 0.86$ and then heated. (**b**) Emulsion was cooled to 7.3°C producing a solidified droplet fraction of $\approx 0.13$ and then heated. In both cases, only the connected solid droplets coalesce during melting while their neighboring liquid drops remain intact. Initial droplet size is 40 μm and cooling/heating rate is 1°C/min. Scale bar denotes 100 μm.

Our observations in Fig. 2b reveal the following key features of the mechanism responsible for the catastrophic destabilization of the 2D emulsion. First, due to the hexagonal order each crystalline droplet is forced to contact six neighbors and strikingly all the contacts form liquid bridges during melting generating system-wide connectivity (Fig. 2c). Thus, the geometric constraints in the system eliminate barriers to synchronous initiation of coalescence. Second, the system-wide connectivity ensures that the entire droplet array relaxes into a large mass of dispersed phase. We note that during the relaxation process, triangular-shaped pockets of continuous phase (see inset of Fig. 2b-i and Fig. 2c) originally present between droplets form tiny water droplets resembling phase inversion (see Movie SM1). Thus, the system-wide connectivity eliminates kinetic barriers to emulsion destabilization. This is in marked contrast to disordered emulsions, where coalescence propagation and the associated cluster dynamics, and capillarity-driven restructuring (c.f. Fig. 1) are essential drivers of the destabilization process.

To confirm that system-wide connectivity is essential for the full destabilization of the emulsion, we perturbed the connectivity by pausing the cooling process before all droplets crystallize and then melted the emulsion. As shown in Fig. 3, we considered two cases – one where the solidified droplet fraction is high ($\approx 0.86$, Fig. 3a) and the other where the solidified droplet fraction is low ($\approx 0.13$, Fig. 3b). In both cases, since supercooled liquid droplets are present, the system-wide connectivity between solidified droplets is broken. Upon heating, as expected, we observe only the solidified droplets that are connected with each other to partially coalesce forming larger droplets. Interestingly, given that the supercooled liquid droplets remain intact suggests that for partial coalescence to occur, all droplets involved not only need to be in contact but also need to be partially crystalline. The results of Fig. 3 confirm that system-wide network connectivity is an essential criterion for full-scale destabilization of the emulsion and also offers a route to mitigate destabilization by perturbing network connectivity.

Next, we investigated whether the physicochemical properties of the individual droplets could alter the synchronous coalescence initiation in a fully connected hexagonally ordered emulsion. By keeping the volume fraction fixed at $\approx 0.4$, as shown in Table 1, we explored system conditions where the droplet size was changed from 40 μm to 24 μm, the surfactant was switched from ionic SDS to non-ionic Tween 20, and the crystal content $\xi$ was varied from 0.5 – 1.0. Here, $\xi = 1.0$ indicates that the droplet is fully solid when cooled (as in Fig. 2), which was achieved using a single alkane *n*-hexadecane as the dispersed phase. Crystal contents of $\xi = 0.5$, 0.75 and 0.9 were obtained by using a mixed alkane system of *n*-hexadecane and *n*-tetradecane, where the proportion of each alkane in the dispersed phase corresponds to $\xi$. Since *n*-tetradecane has a $T_m \approx 5$ °C, which is far lower than *n*-hexadecane, during our cooling cycle, tetradecane molecules in the droplet remain liquid, while hexadecane molecules crystallize. Thus, by generating emulsions in which the composition of the alkane mixture is varied we were able to control the crystal content of individual droplets in the 2D ordered emulsion.

To compare the thermal stability of the different emulsions, we defined the coalescence fraction $n_c = 1 - N_f/N_i$ where $N_f$ is the number of drops remaining after melting and $N_i$ is the initial number of drops in the array. For example, for the emulsion shown in Fig. 2b, $N_i = 600$ and $N_f = 1$ (considering the one large drop), the coalescence fraction $n_c$ approaches unity. Using this measure, we characterized the extent of destabilization in the different emulsions we tested. Surprisingly, as summarized in Table 1, *n*-hexadecane emulsions with two different droplet sizes and surfactants undergo full destabilization which yields $n_c = 1$. Synchronous initiation of coalescence also occurs in these emulsions suggesting that the emulsion destabilization is more sensitive to droplet contact geometry, *i.e.*, number and orientation of contacts than droplet-specific parameters such as size and surfactant.

**Table 1. Influence of physicochemical properties of droplets on the destabilization of 2D hexagonally ordered emulsions.**

| Diameter (μm) | Surfactant | Crystal content | Volume fraction | Coalescence fraction |
|---|---|---|---|---|
| 40 ± 2 | Tween20 | 1 | 0.42 ± 0.04 | 1 |
| 24 ± 2 | SDS | 1 | 0.45 ± 0.05 | 1 |
| 40 ± 2 | SDS | 1 | 0.41 ± 0.04 | 1 |
| 40 ± 2 | SDS | 0.9 | 0.41 ± 0.04 | 1 |
| 40 ± 2 | SDS | 0.75 | 0.38 ± 0.06 | 0.79 ± 0.09 |
| 40 ± 2 | SDS | 0.5 | 0.42 ± 0.05 | 0.68 ± 0.04 |

In emulsions with droplets having different crystal content, we find that full destabilization occurs when the droplets are 100% or 90% crystalline, i.e. for emulsions with $\xi = 1.0$ and 0.9. However, for emulsions with droplets having 50% and 75% crystal content, the coalescence fraction is 0.68 and 0.79 respectively suggesting that these emulsions did not undergo catastrophic thermal destabilization despite the hexagonal order. Thus, unlike droplet size and surfactant type, crystal content of individual droplets can limit the degree to which emulsions destabilize.

To understand the reason for the lower degree of destabilization in emulsions with droplets having 50% and 75% crystal content, we investigated the extent of synchronous coalescence initiation in these two systems. For example, as shown in Fig. 4a, for the emulsion with $\xi = 0.5$ all droplets are intact at 2.1 °C, however, at 8.2 °C some droplets have fused with their neighbors by coalescence initiation, while other droplets remain intact in the system (highlighted by red circles in the inset of Fig.4a). We also observe similar behavior for the emulsion with $\xi = 0.75$. These observations suggest that in both these systems synchronous coalescence initiation did not occur leaving some intact droplets. To characterize the level of asynchronicity $n_a$ we determine the number of intact droplets at the end of melting step and plot it as a fraction of the initial number of droplets in the array. As shown in Fig. 4c, $n_a = 0.24$ and 0.19 for the emulsion with $\xi = 0.5$ and 0.75 respectively; while for emulsions with $\xi = 0.9$ and 1.0, $n_a$ is close to zero. Thus, crystal content of the droplet directly controls asynchronicity of coalescence initiation in the ordered array.

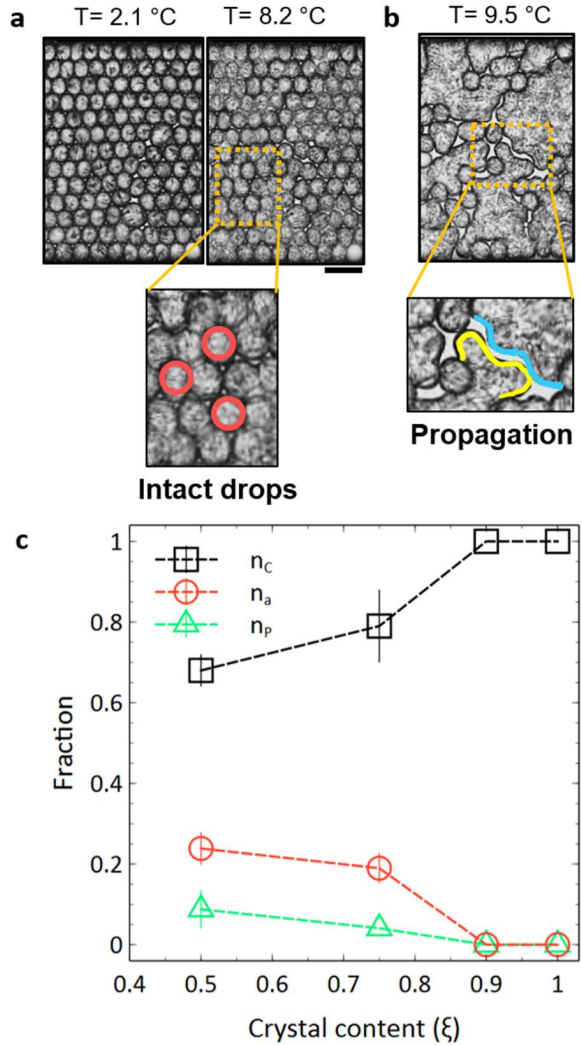

**Fig. 4. Effect of crystal content of drops on the destabilization of hexagonally packed emulsions.** (**a**) Image sequence showing the destabilization process in an emulsion with droplets having 50% crystallinity. Intact droplets survive the melting process. The inset highlights three intact drops (red circles), showing curbing of synchronous coalescence initiation. (**b**) Droplet clusters form that merge during coalescence propagation. The inset highlights merging of the interface of two droplet clusters. Scale bar denotes 100 μm. (**c**) Plots of coalescence fraction ($n_c$), level of asynchronicity ($n_a$) and extent of coalescence propagation ($n_p$). Error bar is standard deviation from two trials.

Accompanying the asynchronicity in coalescence initiation, we observe the emergence of coalescence propagation in the emulsions with lower crystal content. An instance of coalescence propagation is highlighted in Fig. 4b, by the interfaces (yellow and blue, T=9.5 °C) of two droplet clusters that merge. We quantify the coalescence propagation by visually tracking how many fusion events, *i.e.* the number of times the interfaces of two coalescing droplet clusters merge during the destabilization process. These counts of propagation events are normalized by the initial number of droplets in the array and plotted as $n_p$ in Fig. 4c. We find that the propagation events are higher for the emulsions with droplets having lower crystal content. The reason for this is that the asynchronicity in coalescence initiation is higher in emulsions with lower crystal content, causing to form more droplet clusters and therefore more chance for propagation events to occur. Stated differently, when asynchronicity is nearly zero, there is only one large cluster and therefore negligible chance for propagation events to occur, as we show in Fig. 2. Taken together, our results in Fig. 4 indicate that the coalescence fraction at the end of the melt cycle is a result of degree of asynchronicity and frequency of propagation events.

In summary, our work highlights that droplet contact geometry is vital for large-scale emulsion destabilization. We find that the close-packed hexagonal order in fully crystalline droplets promotes synchronous initiation of coalescence and emergence of system-wide connectivity, which virtually eliminates geometric and kinetic barriers to emulsion destabilization, which is in direct contrast to disordered emulsions. The synchronous initiation of coalescence and system-wide connectivity still persists when droplet size is varied or surfactant is changed, as long as the droplets are fully solidified. Remarkably, decreasing the droplet crystal content reduces the extent of emulsion destabilization by curbing synchronous coalescence initiation and introducing coalescence propagation into the system which presents a kinetic barrier for total destabilization of the emulsion.

In the broader context, the mechanism of catastrophic thermal destabilization of ordered 2D emulsions elucidated here can also play a role in phase separation of bulk oil-in-water emulsions subjected to temperature fluctuations. In bulk emulsions, the droplets are dispersed in three dimensions, however, the oil droplets being lighter might cream and form close-packed layers. Synchronous initiation of coalescence along with system-wide connectivity can occur in these close-packed layers resulting in bulk destabilization of the emulsion. Finally, even though we have focused on emulsion destabilization due to phase transitions, a similar phenomenon of sudden destruction of concentrated emulsions through coalescence has been reported in purely liquid 2D emulsions undergoing shear [34]. In this case, destabilization occurs via propagation of coalescence between droplets by a unique separation-driven coalescence mechanism [35] and triggering of multiple coalescence events [34]. In contrast, in our hexagonally ordered array of melting drops, the coalescence initiation occurs spontaneously between every drop and its six neighbors, leading to system-wide connectivity.

**Acknowledgements**

The authors acknowledge the financial support of National Science Foundation (CAREER: 1150836) and the Jack Maddox Distinguished Engineering Chair Professorship in Sustainable Energy, sponsored by the J.F Maddox Foundation. We are grateful to Patrick Spicer for useful discussions.